# Open-tunneled oxides as intercalation host for multivalent ion (Ca and Al) batteries: A DFT study

Joy Datta[1], Nikhil Koratkar[2], and Dibakar Datta[1,*]

[1]Department of Mechanical and Industrial Engineering, New Jersey Institute of Technology (NJIT), Newark, NJ 07052, USA

[2]Department of Mechanical, Aerospace, and Nuclear Engineering, Rensselaer Polytechnic Institute, Troy, New York, 12180, USA

Corresponding author: Dibakar Datta (dibakar.datta@njit.edu)

**Abstract**

Lithium-ion batteries (LIBs) are ubiquitous in everyday applications. However, Lithium (Li) is a limited resource on the planet and is therefore not sustainable. As an alternative to lithium, earth-abundant and cheaper multivalent metals such as aluminum (Al) and calcium (Ca) have been actively researched in battery systems. However, finding suitable intercalation hosts for multivalent-ion batteries is urgently needed. Open-tunneled oxides are a particular category of microparticles distinguished by the presence of integrated one-dimensional channels or nanopores. This work focuses on two promising open-tunnel oxides, viz: Niobium Tungsten Oxide (NTO) and Molybdenum Vanadium Oxide (MoVO). We find that the MoVO structure can adsorb greater numbers of multivalent ions than NTO due to its larger surface area and different shapes. The MoVO structure can adsorb Ca, Li, and Al ions with adsorption potential at around 4 to 5 eV. However, the adsorption potential for hexagonal channels of Al ion drops to 1.73 eV because of less channel area. NTO structure has an insertion/adsorption potential of 4.4 eV, 3.4 eV, and 0.9 eV for one Li, Ca, and Al, respectively. In general, Ca ion is more adsorbable than Al ion in both MoVO and NTO structures. Bader charge analysis and charge density plot reveals the role of charge transfer and ion size on the insertion of multivalent ions such as Ca and Al into MoVO and NTO systems. Our results provide general guidelines to explore other multivalent ions for battery applications.

1. INTRODUCTION

Rechargeable (i.e., secondary) batteries represent a critical enabling technology for electrification of automobile as well as storage of energy from renewables such as wind and solar[1-3]. The sizing of the active material particle used in the anode and cathode plays a crucial role in affecting battery performance.[4,5] Traditionally, microparticles have been used in the industry due to their higher volumetric energy density, high mass loading, better scalability, and lower cost[6]. However, in terms of mechanics, thermodynamics, and kinetics, microparticles face several challenges which nanoparticles can solve[3]. Nanoparticles have improved cycle stability over microparticles. Smaller size nanoparticles can intercalate more uniformly compared to microparticles. This improves fracture toughness and fatigue life for electrode materials[7–9]. Fast charging ability is another superior performance attribute associated with nanoparticles. The small particle size of nanoparticles drastically reduces diffusion length, enabling faster charging and discharging ability[10,11]. Moreover, the limited compositional range makes it less favorable for distinct phases to co-exist in nanostructures. As a result, phase transitions are much quicker for nanostructures to release the excess free energy generated from a lattice mismatch and high surface area[12–14].

Nanostructures also have their distinct set of disadvantages. For example, the nanostructures' large surface makes results in extensive electrolyte decomposition. As a result, the Solid Electrolyte Interphase (SEI) layer forms during the early stage of the battery cycle, causing low first-cycle coulombic efficiency[15–17]. With nanostructures, it is also challenging to achieve an industrial standard (20–30 mg/cm$^2$) high mass loading.[18,19] Nanostructure-based batteries also suffer from low volumetric capacity, which is a major limitation in stationary and grid storage applications[20,21]. The synthesis cost of nanostructures is also higher and produces significant chemical waste in the manufacturing process[22–25].

Multiscale particles (MP) have nanoscale features or attributes in microscale particles. These particles have the characteristics of both micro and nanostructures[3] and could potentially offer the best of both worlds. MP can be synthesized as Engineered Multiscale Particles (E-MP) or as Multiscale Particles with Natural nano-porosity (N-MP). E-MP can be synthesized in several ways. Controlled assembly of nanoparticles to form microparticles or engineering of nanopores in microparticles by etching techniques are some of the common synthesis methods used to produce E-MPs[26–31]. In general, manufacturing cost and scalability are significant challenges for E-MPs.[3, 32] On the other hand, open-tunnel oxide based micro-particles like Niobium Tungsten Oxide (NTO) and Molybdenum Vanadium Oxide (MoVO) are typical N-MPs and come with naturally formed nanoscale channels. They have a superior ability to offer fast ion diffusion[33,34]. There are several family members in NTO structures like $Nb_{12}WO_{33}$, $Nb_{16}W_5O_{55}$, and

$Nb_{18}W_{16}O_{93}$ [35,36]. For example, $Nb_{12}WO_{33}$ has a Wadsley-Roth type crystallographic shear structure. It consists of a (3×4) size of $MO_6$ (M=Nb, W) octahedral[36]. These octahedral blocks share edges around the corner, forming open channel-like structures, shown in Fig. 1a. MVO structures (Fig. 1b, c, d) can have different polymorphs - orthorhombic ($MoV_2O_8$), trigonal ($MoV_3O_6$), and tetragonal ($MoVO_5$). Moreover, they have other tunnels, e.g., hexagonal, heptagonal, pentagonal, rectangular, etc[33].

The cost of rare earth metal Li is increasing rapidly[37,38]. One of the alternatives is to use multivalent ions like Ca and Al[39,40]. Since these ions are multivalent, they can offer similar energy density to Li, despite their greater atomic mass and volume. Ca and Al metal are also earth abundant and low-cost.[41–43] Our objective in this work is to study intercalation of multivalent ions such as Ca and Al into N-MPs such as NTO and MoVO. In terms of prior art, Kocer et al.[35] have investigated Li insertion on three types of Wadsley–Roth crystallographic shear structures named $Nb_{12}WO_{33}$, $Nb_{14}W_3O_{44}$, and $Nb_{16}W_5O_{55}$. Density Functional Theory (DFT) was adopted to find 3 active Li insertion sites: vertical window, horizontal window, and pocket in $Nb_{12}WO_{33}$ structure. Structural evolution was calculated for $Li_xNb_{12}WO_{33}$ by varying Li concentration between $x$=1 and $x$=13. High Li concentrated structure removes $MO_6$(M=Nb, W) octahedral sites, which causes lattice contraction. This lattice contraction helps to mitigate volume expansion and improves long-term cycle life.

In contrast to the above work with Li, there are no reports to date on insertion of multivalent ions such as Al and Ca into NTO or MoVO systems. In this work, we considered trigonal ($MoV_3O_6$) structure consisting of hexagonal, heptagonal, and triangular channels with different sizes. Ca, Al and Li are inserted into these MoVO structures. We measured the adsorption potential of Ca, Li, and Al into those channels. We extended the same study to NTO ($Nb_{12}WO_{33}$) systems and analyzed the adsorption ability of NTO vs. MoVO structures and focused on the effect of multivalent ion size and charge density on insertion potentials. Our overall conclusion is that MoVO structures are more suitable for multivalent ion insertion than NTO structures. Also, our findings elucidate the role of ion size and charge density on intercalation into MoVO and NTO structures.

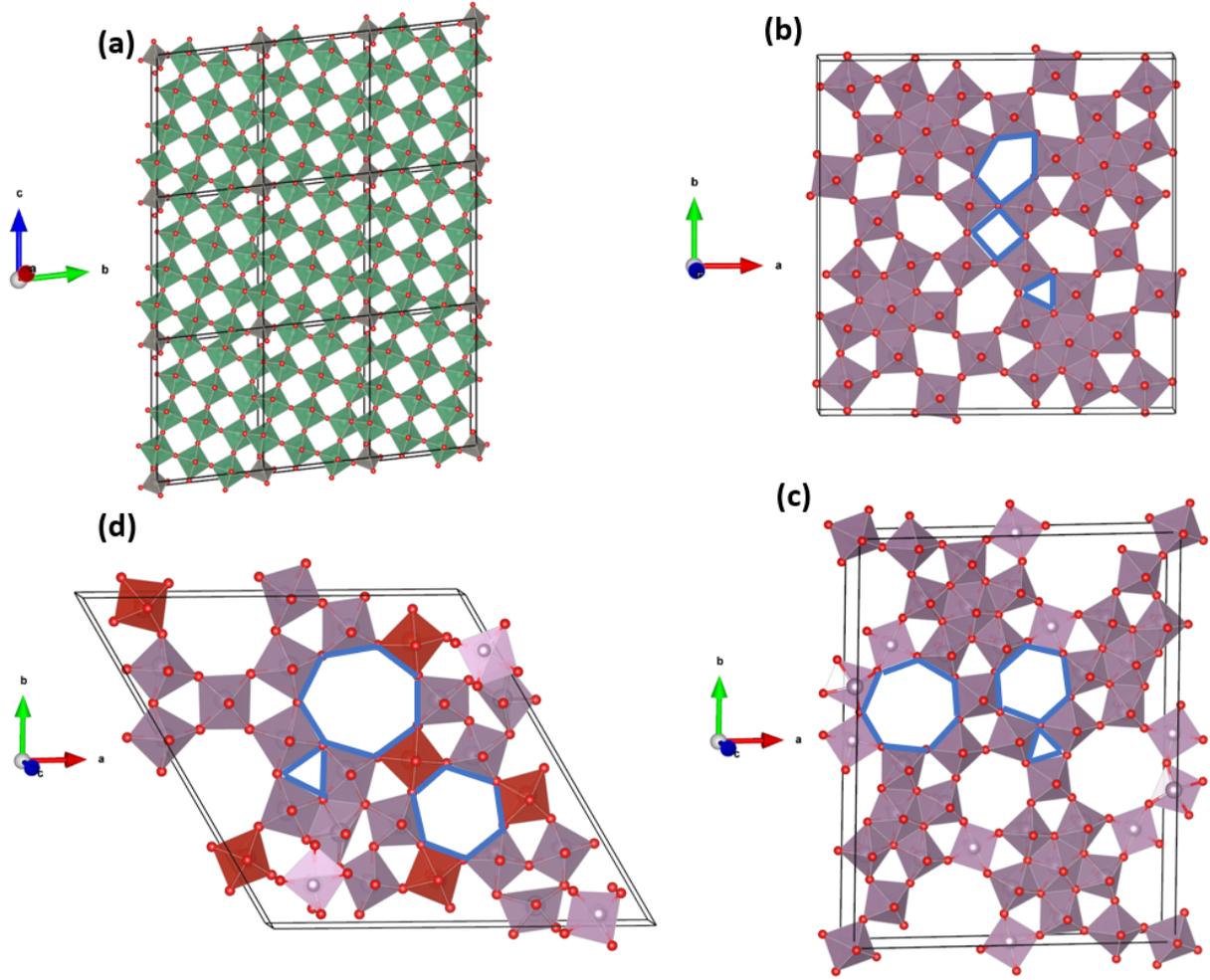

**Figure 1: Crystal structure of N-MP.** (a) $Nb_{12}WO_{33}$ structure: Block size is denoted as $(3\times4)ReO_3$ structure block. Every (3×4) block is joined by a Niobium atom at the tetrahedral junctions and octahedral positions. These crystallographic edge-sharing structures generate channels. (b) Tetragonal ($MoVO_5$) structure consists of pentagonal, rectangular, and triangular channels. (c) Orthorhombic ($MoV_2O_8$) structure consists of heptagonal, hexagonal, and triangular channels. (d) Trigonal ($MoV_3O_6$) structure has a similar type of channel to the orthorhombic but the atom numbers are different. All the channels in (b-d) is shown as blue. Channel dimension of tetragonal MoVO (shown in b) is smaller than orthorhombic and trigonal polymorphs (shown in c, d).

## 2. METHODS

We have used DFT for structural optimization. The Vienna Ab initio Simulation Package was used for the DFT calculations (VASP)[51] with the help of the projector augmented wave (PAW)[52] method. The Perdew−Burke−Ernzerhof (PBE)[53] form of the generalized gradient approximation (GGA) represents the exchange-correlation function. The energy cut-off is 520 eV for representing the plane wave basis set. The k-point grid speaks for the sampling of Brillouin zones. For Molybdenum Vanadium Oxide (MoVO), the k-point is 4×4×1, and for Niobium Tungsten Oxide (NTO) structure, it is 1×4×5. To relax the MoVO and NTO structures, we selected a force tolerance of 0.02 eV/Å and an energy-stopping criterion of $10^{-6}$ eV/Å.

The adsorption potential (V) has been calculated as -

$$V = \frac{\Delta G}{n_f} \quad (1)$$

Where n is the concentration of Ca/Li/Al ion. The Gibbs free energy $\Delta G$ is defined by -

$$\Delta G = \Delta E_f + P\Delta V_f - T\Delta S_f \quad (2)$$

In equation (2), at room temperature, $P\Delta V_f$ and $T\Delta S_f$ are very negligible compared to $\Delta E_f$ [4]. Formation energy $\Delta E_f$ can be computed using this equation-

$$\Delta E_f = \Delta E_{X_nG} - (nE_x + E_g) \quad (3)$$

Where $E_{X_nG}$ is the total energy of Ca/Li/Al intercalated MoVO/NTO structure, $E_x$ represents total energy of single Ca/Li/Al ion, $E_g$ is the total energy of MoVO/NTO structures. For this work, equilibrium energy of Ca and Li are -1.980 eV[54] and -1.8978 eV[55], respectively. We obtained Al equilibrium energy as -3.45 eV by performing DFT calculations.

## 3. RESULTS AND DISCUSSION

We inserted Ca into trigonal MoVO structures into three different types of channels named heptagonal, hexagonal, and triangular shape[5] because they contain 7, 6, and 3-member rings, respectively. Finding the most stable ion position in any specific tunnel is crucial. For this work, we inserted Ca in several places in the hexagonal and heptagonal tunnels and considered the most stable position by comparing the energy of the optimized structure. Heptagonal and hexagonal have a 5-6 Å channel dimension, which is five times bigger than the Ca[33]. Therefore, they can easily intercalate Ca into their channels which is reflected in Figure 2(b-f). As the triangular tunnel has limited space, it cannot accommodate more than 1 Ca. To calculate the difference in adsorption potential more clearly, we inserted 2 Ca into the heptagonal and hexagonal tunnels. For 1 Ca, heptagonal and hexagonal tunnels have almost similar adsorption favorability of 4 eV (Figure 2b, c, g). Next, we studied the insertion of 2 Ca into the heptagonal and hexagonal tunnels. In this case, the adsorption potential for the heptagonal and hexagonal channels is 3.06 eV and 2.07 eV, respectively. Therefore, the heptagonal channel is more favorable for Ca insertion than other channels.

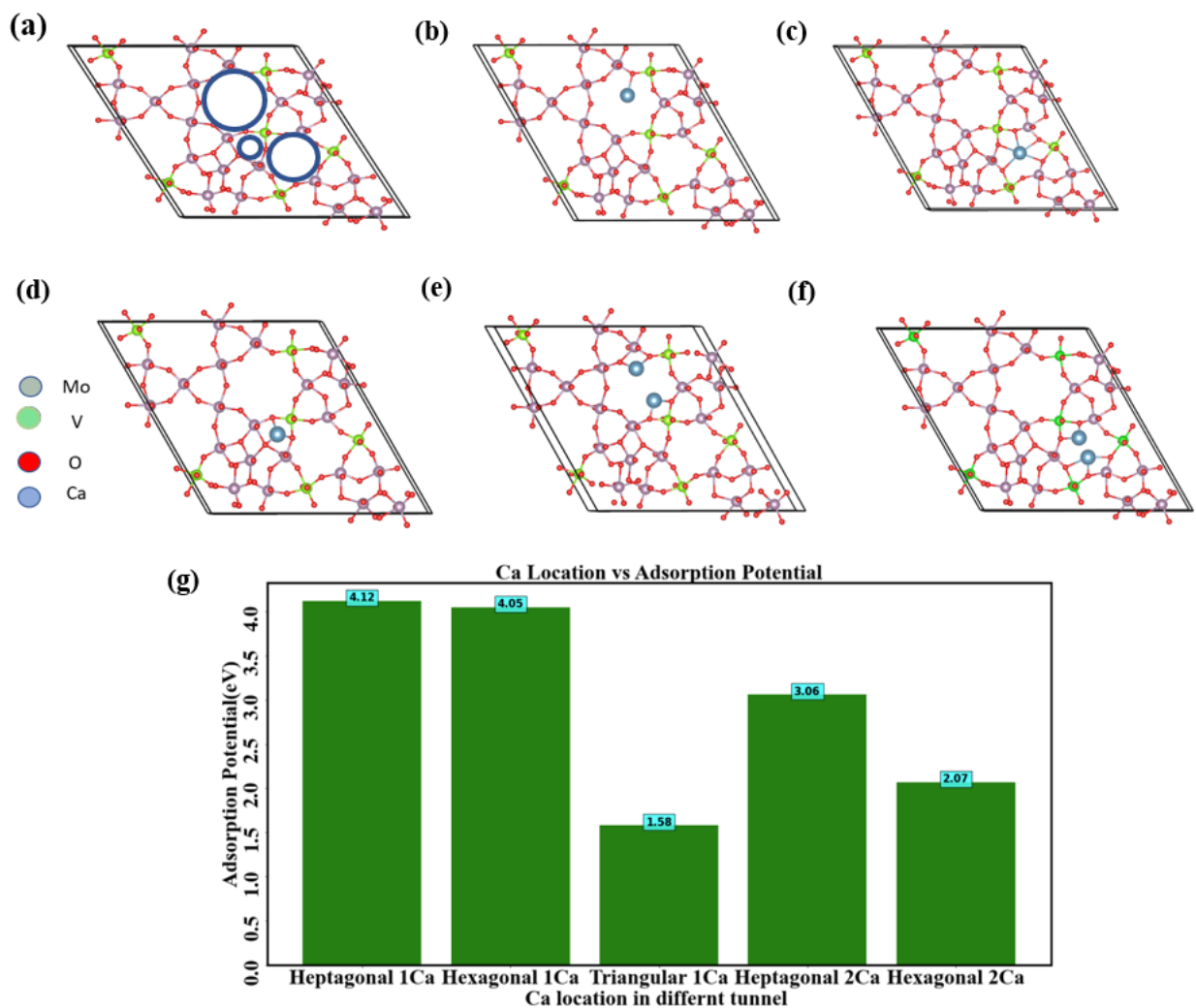

**Figure 2.** Inserting Ca in triangular MoVO structure (a) Trigonal MoVO structure without Ca (b) 1 Ca in heptagonal channel (c) 1 Ca in hexagonal channel (d) 1 Ca in trigonal channel (e) 2 Ca in heptagonal channel (f) 2 Ca in hexagonal channel (g) Ca location w.r.t. adsorption potential. Heptagonal channel is the most favorable location for Ca than any other channel. Triangular channel cannot place 2 Ca due to the small channel dimension.

As Li has smaller charge density (Table 1) and size compared to Ca, we repeated the above study with Li to study role of charge density and particle size. Li inserts into the heptagonal and hexagonal channels by changing locations to find the most optimized structure. Adsorption potential for 1 Li is highest for the heptagonal tunnel (4.73 eV, Figure 3a, f), and lowest for the triangular channel (3.53 eV, Figure 3c, f). Figure 3f shows that Li's adsorption potential is relatively high after inserting 2 Li into the heptagonal and hexagonal channels. The adsorption potential for the heptagonal and hexagonal channels is 4.22 eV and 3.46 eV, respectively. Thus particle size plays a significant advantage for 2 Li to gain high adsorption potential (Figure 3f).

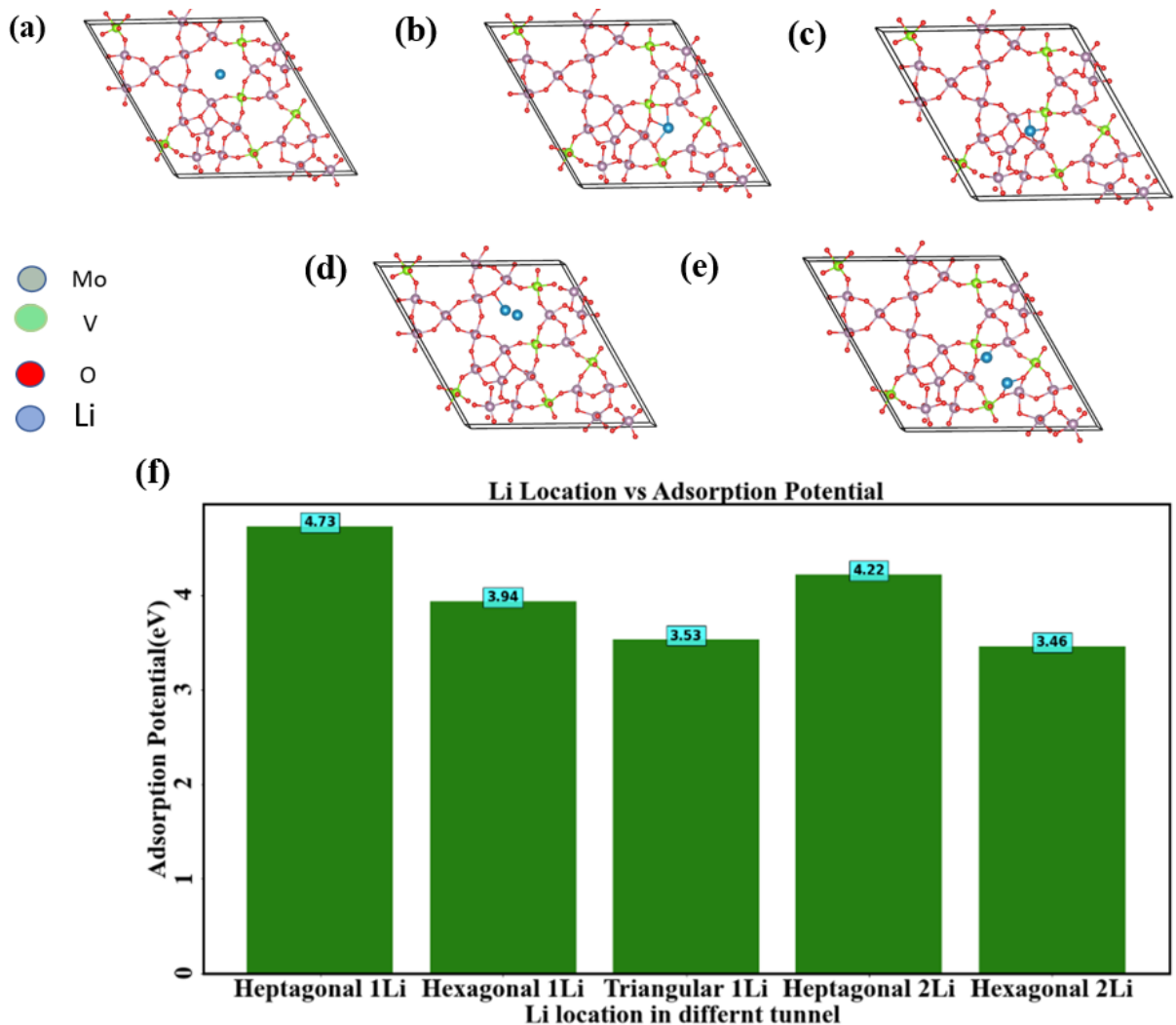

**Figure 3.** Inserting Li in triangular MoVO structure (a) 1 Li in heptagonal channel (b) 1 Li in hexagonal channel (c) 1 Li in trigonal channel (d) 2 Li in heptagonal channel (e) 2 Li in hexagonal channel (f) Li location w.r.t adsorption potential. The adsorption favorability of Li ion into channels varies in terms of channel dimension.

Al is inserted following the same intercalation process to compare the ability of those channels. For triangular channels (Figure 4c), Al cannot be inserted as the adsorption potential at this channel is -1.95 eV (Figure 4f). The highest adsorption favorability of 1 Al is for the heptagonal channel, which is 4.39 eV (Figure 4f). The adsorption potential of the hexagonal tunnel is less than 50% compared to the heptagonal channel (Figure 4f). A bigger channel size favors Al to achieve this high adsorption potential for the heptagonal channel (Figure 4f) as heptagonal channel has 1.5 times more area in comparison with hexagonal channel. Also, for hexagonal channels, the ability to capture 2 Al is challenging because of low adsorption favorability (0.66 eV, Figure 4f).

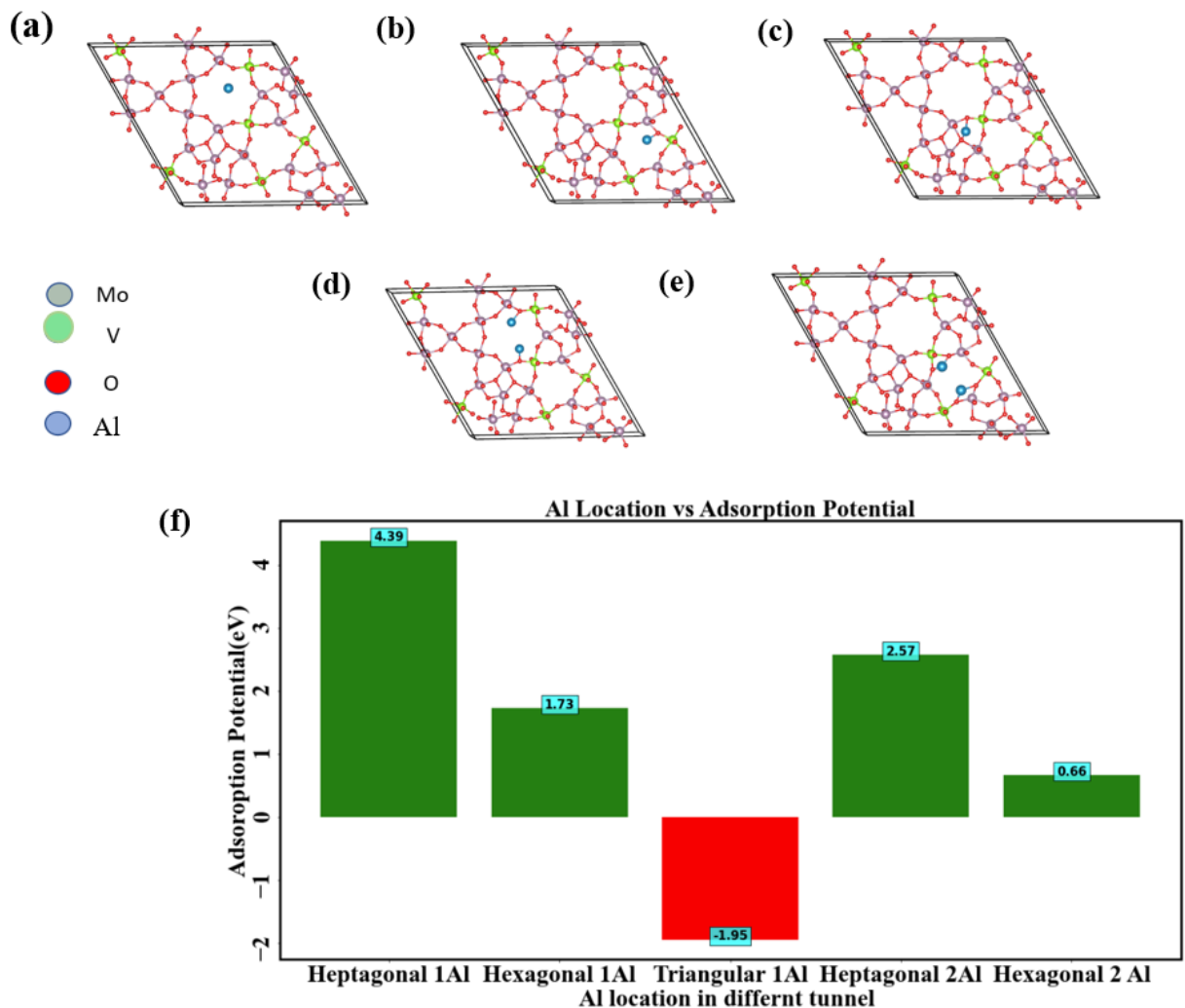

**Figure 4.** Inserting Al in triangular MoVO structure (a) 1 Al in heptagonal channel (b) 1 Al in hexagonal channel (c) 1 Al in trigonal channel (d) 2 Al in heptagonal channel (e) 2 Al in hexagonal channel (f) Al location w.r.t. adsorption potential. Al has the high charge density issue. Hence it has negative adsorption potential for triangular channel (Table 1).

Figure 5 shows the performance comparison regarding the adsorption ability of Ca, Li, and Al. The Li has the highest adsorption ability in any channel shape. Ca has approximately the same intercalation ability as Li, though Ca has larger particle size compared to Li and Al. As the channel dimension gets smaller from heptagonal to triangular, Al shows poor adsorption potential compared to Li and Ca (figure 5). The main reasons behind this is that it possesses the highest charge density (Table 1).

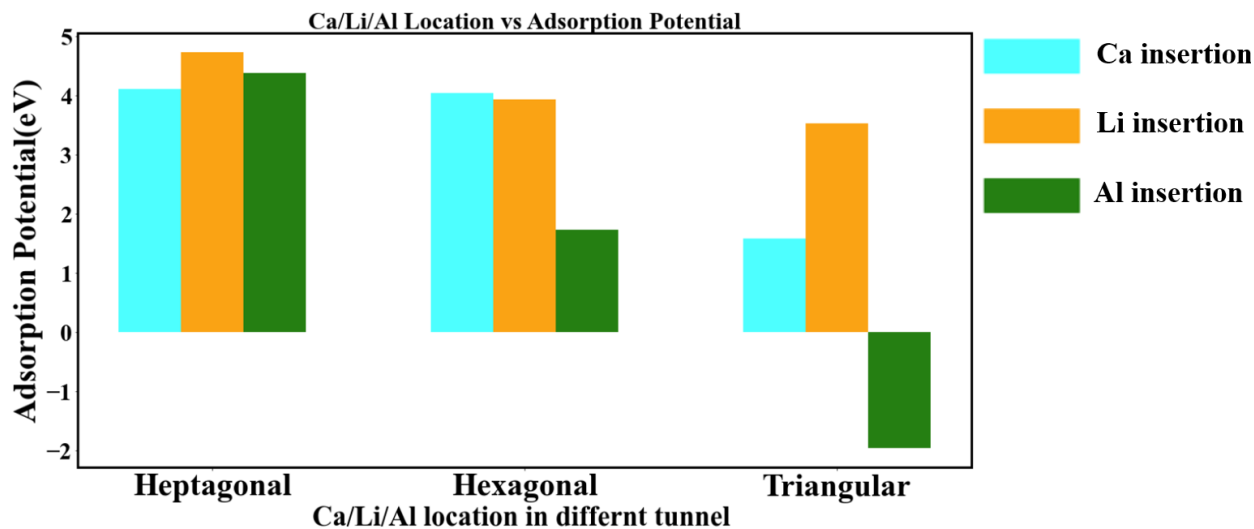

**Figure 5.** Comparison of adsorption potential of 1 Ca/Li/Al w.r.t different tunnel location. For heptagonal channel the adsorption potential sees similar performance among all due its high channel dimension. The charge density effect is obvious at triangular channel.

Another multiscale particle with naturally occurring nanochannels is the NTO-based structure. One of the family members of the NTO structure is $Nb_{12}WO_{33}$ [35,36]. Like MoVO structures, we computed the adsorption potential for Li/Ca/Al insertion into NTO using DFT. Only one rectangular shape channel is present in $Nb_{12}WO_{33}$. We inserted Li/Ca/Al into those channels by varying concentrations (Figure 6a-f). For 1 and 2 Li insertion, the adsorption potential (Figure 6g) is 4.4 eV and 3.44 eV, respectively. The 2 Ca and 2 Al insertion into the channel resulted in the adsorption potential of -0.82 eV and -0.26 eV, respectively (Figure 6g). However, insertion favorability of 2 Ca and 2 Al is positive for MoVO structure because heptagonal and hexagonal channel has 41.67 $Å^2$ and 27.19 $Å^2$, respectively. However, for NTO, the area is only 15.36 $Å^2$. These results indicate that channel area and charge density of cation (Table 2) plays an important factor for adsorption favorability. Our results indicate that inserting multivalent ions into the MoVO structure is more advantageous as compared to NTO.

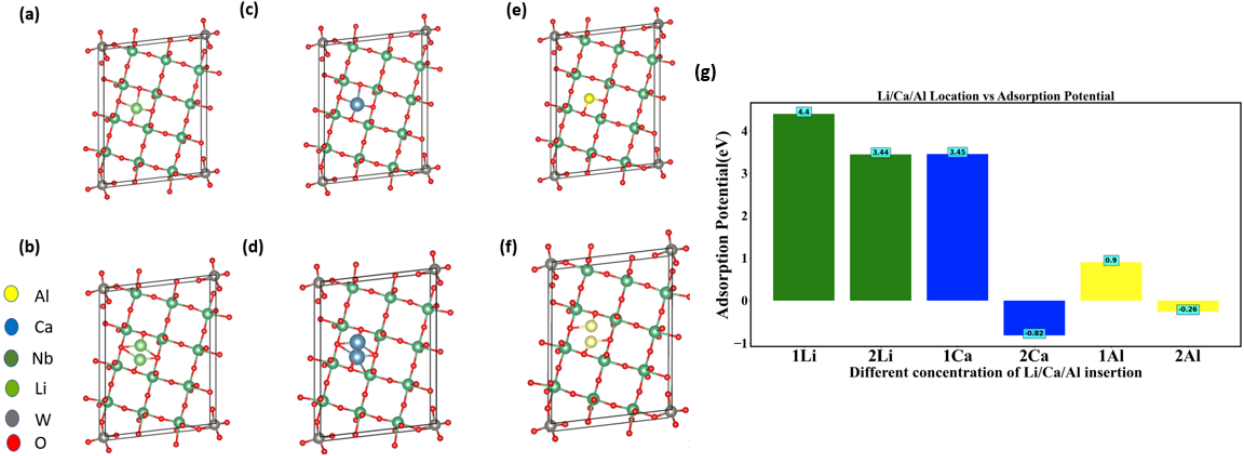

**Figure 6.** Inserting different concentrations of Li/Ca/Al in $Nb_{12}WO_{33}$ structure: (a) 1 Li, (b) 2 Li, (c) 1 Ca, (d) 2 Ca, (e) 1 Al, (f) 2 Al, (g) Different concentration of Li/Ca/Al insertion vs. adsorption potential. Li has the highest adsorption potential due to the lowest particle size and less charge density compared to Ca and Al.

To obtain a detailed idea about the charge distribution inside the MoVO and NTO structures, Bader charge analysis and charge density difference has been calculated[56].

$$\Delta\rho = \rho_{A+B} - (\rho_A + \rho_B) \quad (4)$$

Here, A is MoVO/NTO crystal structure, and B is the cation inserted into this system, $\rho_A$ indicates the density of MoVO/NTO crystal structures and $\rho_B$ is the density of the cation. Bader charge analysis has determined the number of charge transfers between cations and MoVO structures. Table 1 shows that 1 Al ion charge transferred to a triangular channel has a charge transfer of 1.65e, which is higher than other cations. The charge density plot in Figure 7 shows the relationship between charge accumulation and depletion. In Figure 7m, Al ion has the weakest interactions with all cations at these positions. Therefore, the negative adsorption potential, shown in Figures 4 and 5, is due to the lowest ionization of Al ion and MoVO structure at the triangular channel.

Charge transfer for NTO structures is highest for 1 Al ion (Table 2). Consequently, the adsorption potential is, therefore, the lowest for one ion insert (Figure 6g). Concerning two ions insertion, Ca has the highest charge transfer relative to other cations (Table 2). As a result, 2 Ca ions insertion has the least adsorption potential. In general, our charge analyses show inverse correlation between adsorption potential and charge transfer. More charge transfer corresponds to lesser adsorption.

**Table 1:** Calculated charge transfer from cation to the MoVO structures

| | Ion name | Heptagonal 1 ion | Hexagonal 1 ion | Triangular 1 ion | Heptagonal 2 ion | Hexagonal 2 ion |
|---|---|---|---|---|---|---|
| Charge transfer(e) | Ca | 1.08 | 1.57 | 1.42 | 1.26 | 1.24 |
| | Li | 0.55 | 0.69 | 0.36 | 0.35 | 0.46 |
| | Al | 1.28 | 1.044 | 1.65 | 1.1 | 1.45 |

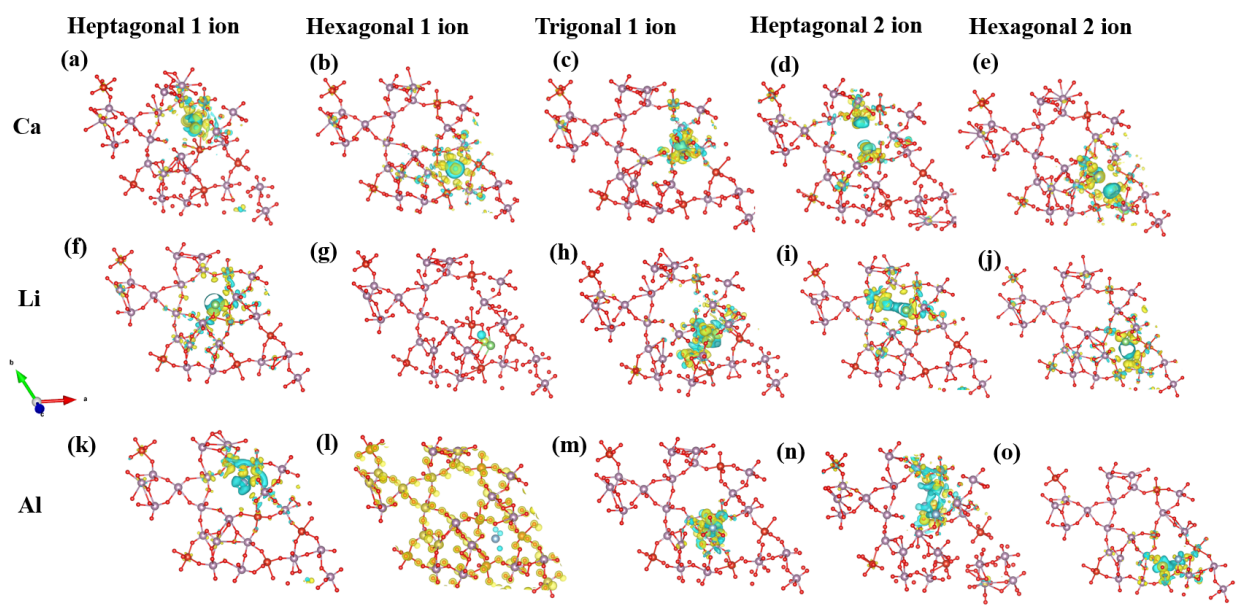

**Figure 7:** Charge density plot of MoVO structure for Ca, Li and Al ion: (a-c) 1 Ca insertion at different channel site, (d, e) 2 Ca insertion at heptagonal and hexagonal channel, (f-g) 1 Li insertion at different channel site, (i,j) 2 Li insertion at heptagonal and hexagonal channel, (k-m) 1 Al insertion at different channel site, (n,o) 2 Al insertion at heptagonal and hexagonal channel. The isosurface level is set at 0.0045 e/ $Å^3$. The yellow and cyan color represents charge accumulation and charge depletion respectively.

Table 2: Calculated charge transfer from cation to the NTO structures

|  |  | 1 ion | 2 ion |
|---|---|---|---|
| Charge transfer(e) | Ca | 1.3667 | 1.2708 |
|  | Li | 0.22245 | 0.2584 |
|  | Al | 1.5308 | 0.96 |

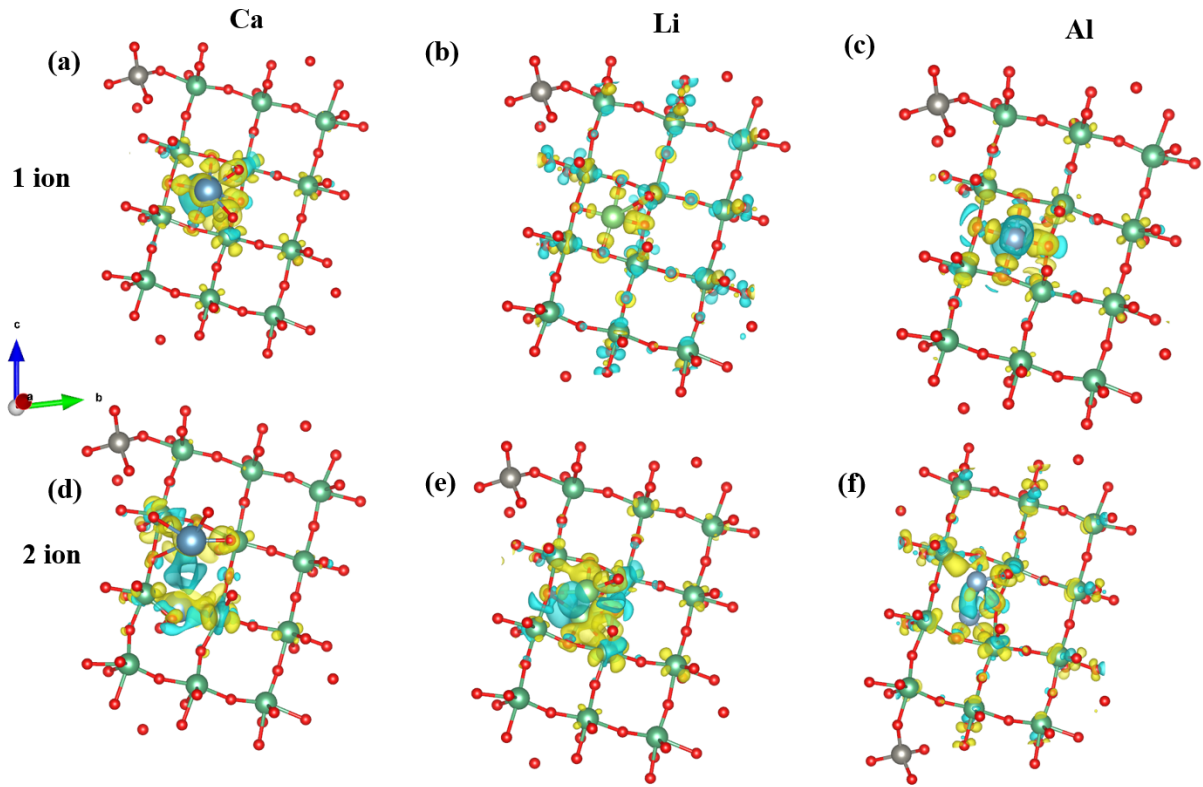

**Figure 8:** Charge density plot of NTO structure for Ca, Li and Al ion: (a-c) 1 Ca/Li/Al insertion, (d, e) 2 Ca/Li/Al insertion. The isosurface level is set at 0.0088 e/Å$^3$. The yellow and cyan color represents charge accumulation and charge depletion respectively.

## 4. CONCLUSIONS

First-principles calculations were performed by inserting mono (Li) and multivalent (Ca, Al) species into multiscale oxide particles with one-dimensional nanochannels such as NTO and MoVO. Comparison of the adsorption potential of Ca, Li, and Al ions shows that the MoVO structure is more favorable to accommodate multivalent ions than NTO. Large channel areas and different shapes of channel availability, like hexagonal, heptagonal, and triangular shapes, make MoVO a superior intercalation host for multivalent ions as compared to NTO structures.

Results obtained from Bader charge analysis and charge density plot reveal that Al ion has the lowest insertion favorability for both MoVO and NTO structures as Al has the highest charge transfer to the MoVO and NTO structures. Ca has adsorption behavior similar to monovalent Li, but it faces challenges in the small channels of NTO structures when 2 Ca ions are inserted. Therefore, channel dimension, ion size, and ion charge density play a crucial role in ions insertion into MoVO and NTO structures. This research can pave the way for the exploration of other candidate open-tunneled oxide materials, for superior performing multivalent-ion batteries.

**AUTHOR INFORMATION**

**Corresponding Author**
Dibakar Datta
Email: dibakar.datta@njit.edu
Phone: +1 973 596 3647

**Author Contributions**
D.D. and N.K. conceived the project. J.D. performed all work and wrote the manuscript with D.D. and N.K. All authors approved the final version of the manuscript.

**CONFLICT OF INTEREST STATEMENT**

The authors have no conflicts of interest to declare. All authors have seen and agree with the contents of the manuscript and there is no financial interest to report. We certify that the submission is original work and is not under review at any other publication.

ACKNOWLEDGEMENT

The work is supported by National Science Foundation (NSF) Electrochemical Systems program (Award Number # 2126180). Authors acknowledge Advanced Cyberinfrastructure Coordination Ecosystem: Service & Support (ACCESS) for the computational facilities (Award Number - DMR180013).

DATA AVAILABILITY

The data reported in this paper is available from the corresponding author upon reasonable request.

CODE AVAILABILITY

The pre- and post-processing codes used in this paper are available from the corresponding author upon reasonable request. Restrictions apply to the availability of the simulation codes, which were used under license for this study.

**References**

(1) Lea, R. Surging Oil and Gas Prices, as Central Banks Due to Meet. *Arbuthnot Bank. Gr.* **2022**, *7*.

(2) Hundekar, P.; Jain, R.; Lakhnot, A. S.; Koratkar, N. Recent Advances in the Mitigation of Dendrites in Lithium-Metal Batteries. *J. Appl. Phys.* **2020**, *128* (1), 10903.

(3) Jain, R.; Lakhnot, A. S.; Bhimani, K.; Sharma, S.; Mahajani, V.; Panchal, R. A.; Kamble, M.; Han, F.; Wang, C.; Koratkar, N. Nanostructuring versus Microstructuring in Battery Electrodes. *Nat. Rev. Mater.* **2022**, *7* (9), 736–746. https://doi.org/10.1038/s41578-022-00454-9.

(4) Chu, S.; Cui, Y.; Liu, N. The Path towards Sustainable Energy. *Nat. Mater.* **2017**, *16* (1), 16–22.

(5) Chu, S.; Majumdar, A. Opportunities and Challenges for a Sustainable Energy Future. *Nature* **2012**, *488* (7411), 294–303.

(6) Whittingham, M. S. History, Evolution, and Future Status of Energy Storage. *Proc. IEEE* **2012**, *100* (Special Centennial Issue), 1518–1534.

(7) Qi, W.; Shapter, J. G.; Wu, Q.; Yin, T.; Gao, G.; Cui, D. Nanostructured Anode Materials for Lithium-Ion Batteries: Principle, Recent Progress and Future Perspectives. *J. Mater. Chem. A* **2017**, *5* (37), 19521–19540.

(8) Graetz, J.; Ahn, C. C.; Yazami, R.; Fultz, B. Highly Reversible Lithium Storage in Nanostructured Silicon. *Electrochem. Solid-State Lett.* **2003**, *6* (9), A194.

(9) Liu, X. H.; Zhong, L.; Huang, S.; Mao, S. X.; Zhu, T.; Huang, J. Y. Size-Dependent Fracture of Silicon Nanoparticles during Lithiation. *ACS Nano* **2012**, *6* (2), 1522–1531.

(10) Tang, Y.; Zhang, Y.; Li, W.; Ma, B.; Chen, X. Rational Material Design for Ultrafast Rechargeable Lithium-Ion Batteries. *Chem. Soc. Rev.* **2015**, *44* (17), 5926–5940.

(11) Tsai, P.-C.; Wen, B.; Wolfman, M.; Choe, M.-J.; Pan, M. S.; Su, L.; Thornton, K.; Cabana, J.; Chiang, Y.-M. Single-Particle Measurements of Electrochemical Kinetics in NMC and NCA Cathodes for Li-Ion Batteries. *Energy Environ. Sci.* **2018**, *11* (4), 860–871.

(12) Madej, E.; La Mantia, F.; Schuhmann, W.; Ventosa, E. Impact of the Specific Surface Area on the Memory Effect in Li-Ion Batteries: The Case of Anatase TiO2. *Adv. Energy Mater.* **2014**, *4* (17), 1400829.

(13) Van der Ven, A.; Wagemaker, M. Effect of Surface Energies and Nano-Particle Size Distribution on Open Circuit Voltage of Li-Electrodes. *Electrochem. commun.* **2009**, *11* (4), 881–884.

(14) Guo, X.; Song, B.; Yu, G.; Wu, X.; Feng, X.; Li, D.; Chen, Y. Size-Dependent Memory Effect of the LiFePO4 Electrode in Li-Ion Batteries. *ACS Appl. Mater. Interfaces* **2018**, *10* (48), 41407–41414.

(15) Xu, K. Electrolytes and Interphases in Li-Ion Batteries and Beyond. *Chem. Rev.* **2014**, *114* (23), 11503–11618.

(16) Keller, C.; Desrues, A.; Karuppiah, S.; Martin, E.; Alper, J. P.; Boismain, F.; Villevieille, C.; Herlin-Boime, N.; Haon, C.; Chenevier, P. Effect of Size and Shape on Electrochemical Performance of Nano-Silicon-Based Lithium Battery. *Nanomaterials* **2021**, *11* (2), 307.

(17) Lai, S. Y.; Knudsen, K. D.; Sejersted, B. T.; Ulvestad, A.; Mæhlen, J. P.; Koposov, A. Y. Silicon Nanoparticle Ensembles for Lithium-Ion Batteries Elucidated by Small-Angle Neutron Scattering. *ACS Appl. Energy Mater.* **2019**, *2* (5), 3220–3227.

(18) Wang, F.; Robert, R.; Chernova, N. A.; Pereira, N.; Omenya, F.; Badway, F.; Hua, X.; Ruotolo, M.; Zhang, R.; Wu, L. Conversion Reaction Mechanisms in Lithium Ion Batteries: Study of the Binary Metal Fluoride Electrodes. *J. Am. Chem. Soc.* **2011**, *133* (46), 18828–18836.

(19) Courtney, I. A.; McKinnon, W. R.; Dahn, J. R. On the Aggregation of Tin in SnO Composite

Glasses Caused by the Reversible Reaction with Lithium. *J. Electrochem. Soc.* **1999**, *146* (1), 59.

(20) Karkar, Z.; Jaouhari, T.; Tranchot, A.; Mazouzi, D.; Guyomard, D.; Lestriez, B.; Roué, L. How Silicon Electrodes Can Be Calendered without Altering Their Mechanical Strength and Cycle Life. *J. Power Sources* **2017**, *371*, 136–147.

(21) Sun, Y.-K.; Oh, S.-M.; Park, H.-K.; Scrosati, B. Micrometer-Sized, Nanoporous, High-Volumetric-Capacity $LiMn_{0.85}Fe_{0.15}PO_4$ Cathode Material for Rechargeable Lithium-Ion Batteries. *Adv. Mater.* **2011**, *23*.

(22) Jung, S.-K.; Hwang, I.; Chang, D.; Park, K.-Y.; Kim, S. J.; Seong, W. M.; Eum, D.; Park, J.; Kim, B.; Kim, J. Nanoscale Phenomena in Lithium-Ion Batteries. *Chem. Rev.* **2019**, *120* (14), 6684–6737.

(23) Berckmans, G.; Messagie, M.; Smekens, J.; Omar, N.; Vanhaverbeke, L.; Van Mierlo, J. Cost Projection of State of the Art Lithium-Ion Batteries for Electric Vehicles up to 2030. *Energies* **2017**, *10* (9), 1314.

(24) Fan, X.; Zhu, Y.; Luo, C.; Suo, L.; Lin, Y.; Gao, T.; Xu, K.; Wang, C. Pomegranate-Structured Conversion-Reaction Cathode with a Built-in Li Source for High-Energy Li-Ion Batteries. *ACS Nano* **2016**, *10* (5), 5567–5577.

(25) Hsu, K.-F.; Tsay, S.-Y.; Hwang, B.-J. Synthesis and Characterization of Nano-Sized $LiFePO_4$ Cathode Materials Prepared by a Citric Acid-Based Sol–Gel Route. *J. Mater. Chem.* **2004**, *14* (17), 2690–2695.

(26) Lou, X. W.; Archer, L. A.; Yang, Z. Hollow Micro-/Nanostructures: Synthesis and Applications. *Adv. Mater.* **2008**, *20* (21), 3987–4019.

(27) Xu, Y.; Zhu, K.; Liu, P.; Wang, J.; Yan, K.; Liu, J.; Zhang, J.; Li, J.; Yao, Z. Controllable Synthesis of 3D $Fe_3O_4$ Micro-Cubes as Anode Materials for Lithium Ion Batteries. *CrystEngComm* **2019**, *21* (34), 5050–5058.

(28) He, H.; Fu, C.; An, Y.; Feng, J.; Xiao, J. Biofunctional Hollow γ-$MnO_2$ Microspheres by a One-Pot Collagen-Templated Biomineralization Route and Their Applications in Lithium Batteries. *RSC Adv.* **2021**, *11* (59), 37040–37048.

(29) Zhang, G.; Yu, L.; Wu, H. Bin; Hoster, H. E.; Lou, X. W. Formation of $ZnMn_2O_4$ Ball-in-ball Hollow Microspheres as a High-performance Anode for Lithium-Ion Batteries. *Adv. Mater.* **2012**, *24* (34), 4609–4613.

(30) Pan, A.; Wu, H. Bin; Yu, L.; Lou, X. W. Template-free Synthesis of VO2 Hollow Microspheres with Various Interiors and Their Conversion into V2O5 for Lithium-ion Batteries. *Angew. Chemie* **2013**, *125* (8), 2282–2286.

(31) Partheeban, T.; Sasidharan, M. Template-Free Synthesis of LiV3O8 Hollow Microspheres as Positive Electrode for Li-Ion Batteries. *J. Mater. Sci.* **2020**, *55* (5), 2155–2165.

(32) Wang, J.; Zhou, H.; Nanda, J.; Braun, P. V. Three-Dimensionally Mesostructured Fe2O3 Electrodes with Good Rate Performance and Reduced Voltage Hysteresis. *Chem. Mater.* **2015**, *27* (8), 2803–2811.

(33) Lakhnot, A. S.; Bhimani, K.; Mahajani, V.; Panchal, R. A.; Sharma, S.; Koratkar, N. Reversible and Rapid Calcium Intercalation into Molybdenum Vanadium Oxides. *Proc. Natl. Acad. Sci. U. S. A.* **2022**, *119* (30). https://doi.org/10.1073/pnas.2205762119.

(34) Lakhnot, A. S.; Gupta, T.; Singh, Y.; Hundekar, P.; Jain, R.; Han, F.; Koratkar, N. Aqueous Lithium-Ion Batteries with Niobium Tungsten Oxide Anodes for Superior Volumetric and Rate Capability. *Energy Storage Mater.* **2020**, *27*, 506–513. https://doi.org/10.1016/j.ensm.2019.12.012.

(35) Koçer, C. P.; Griffith, K. J.; Grey, C. P.; Morris, A. J. Cation Disorder and Lithium Insertion Mechanism of Wadsley-Roth Crystallographic Shear Phases from First Principles. *J. Am. Chem. Soc.* **2019**, *141* (38), 15121–15134. https://doi.org/10.1021/jacs.9b06316.

(36) Cava, R. J.; Murphy, D. W.; Zahurak, S. M. Lithium Insertion in Wadsley-Roth Phases Based on Niobium Oxide. *J. Electrochem. Soc.* **1983**, *130* (12), 2345–2351. https://doi.org/10.1149/1.2119583.

(37) Lee, K. T.; Jeong, S.; Cho, J. Roles of Surface Chemistry on Safety and Electrochemistry in Lithium Ion Batteries. *Acc. Chem. Res.* **2013**, *46* (5), 1161–1170.

(38) Komaba, S.; Murata, W.; Ishikawa, T.; Yabuuchi, N.; Ozeki, T.; Nakayama, T.; Ogata, A.; Gotoh, K.; Fujiwara, K. Electrochemical Na Insertion and Solid Electrolyte Interphase for Hard-carbon Electrodes and Application to Na-Ion Batteries. *Adv. Funct. Mater.* **2011**, *21* (20), 3859–3867.

(39) Elia, G. A.; Marquardt, K.; Hoeppner, K.; Fantini, S.; Lin, R.; Knipping, E.; Peters, W.; Drillet, J.; Passerini, S.; Hahn, R. An Overview and Future Perspectives of Aluminum Batteries. *Adv. Mater.* **2016**, *28* (35), 7564–7579.

(40) Gummow, R. J.; Vamvounis, G.; Kannan, M. B.; He, Y. Calcium-ion Batteries: Current State-of-

the-art and Future Perspectives. *Adv. Mater.* **2018**, *30* (39), 1801702.

(41) Gheytani, S.; Liang, Y.; Wu, F.; Jing, Y.; Dong, H.; Rao, K. K.; Chi, X.; Fang, F.; Yao, Y. An Aqueous Ca-Ion Battery. *Adv. Sci.* **2017**, *4* (12), 1–7. https://doi.org/10.1002/advs.201700465.

(42) Tojo, T.; Sugiura, Y.; Inada, R.; Sakurai, Y. Reversible Calcium Ion Batteries Using a Dehydrated Prussian Blue Analogue Cathode. *Electrochim. Acta* **2016**, *207*, 22–27.

(43) Smeu, M.; Hossain, M. S.; Wang, Z.; Timoshevskii, V.; Bevan, K. H.; Zaghib, K. Theoretical Investigation of Chevrel Phase Materials for Cathodes Accommodating Ca2+ Ions. *J. Power Sources* **2016**, *306*, 431–436. https://doi.org/10.1016/j.jpowsour.2015.12.009.

(44) Carrasco, J. Role of van Der Waals Forces in Thermodynamics and Kinetics of Layered Transition Metal Oxide Electrodes: Alkali and Alkaline-Earth Ion Insertion into V2O5. *J. Phys. Chem. C* **2014**, *118* (34), 19599–19607. https://doi.org/10.1021/jp505821w.

(45) Evans, H. A.; Wu, Y.; Seshadri, R.; Cheetham, A. K. Perovskite-Related ReO3-Type Structures. *Nat. Rev. Mater.* **2020**, *5* (3), 196–213. https://doi.org/10.1038/s41578-019-0160-x.

(46) Deringer, V. L. Modelling and Understanding Battery Materials with Machine-Learning-Driven Atomistic Simulations Journal of Physics : Energy OPEN ACCESS Modelling and Understanding Battery Materials with Machine-Learning-Driven Atomistic Simulations. *J. Phys. Energy* **2020**, *2*, 041003-1–11.

(47) He, X.; Zhu, Y.; Epstein, A.; Mo, Y. Statistical Variances of Diffusional Properties from Ab Initio Molecular Dynamics Simulations. *npj Comput. Mater.* **2018**, *4* (1). https://doi.org/10.1038/s41524-018-0074-y.

(48) Deringer, V. L.; Bernstein, N.; Bartók, A. P.; Cliffe, M. J.; Kerber, R. N.; Marbella, L. E.; Grey, C. P.; Elliott, S. R.; Csányi, G. Realistic Atomistic Structure of Amorphous Silicon from Machine-Learning-Driven Molecular Dynamics. *J. Phys. Chem. Lett.* **2018**, *9* (11), 2879–2885.

(49) Wang, Q.; Zhang, G.; Li, Y.; Hong, Z.; Wang, D.; Shi, S. Application of Phase-Field Method in Rechargeable Batteries. *npj Comput. Mater.* **2020**, *6* (1), 1–8.

(50) Bower, A. F.; Guduru, P. R. A Simple Finite Element Model of Diffusion, Finite Deformation, Plasticity and Fracture in Lithium Ion Insertion Electrode Materials. *Model. Simul. Mater. Sci. Eng.* **2012**, *20* (4), 45004.

(51) Kresse, G.; Furthmüller, J. Efficient Iterative Schemes for Ab Initio Total-Energy Calculations

Using a Plane-Wave Basis Set. *Phys. Rev. B* **1996**, *54* (16), 11169.

(52) Kresse, G.; Joubert, D. From Ultrasoft Pseudopotentials to the Projector Augmented-Wave Method. *Phys. Rev. b* **1999**, *59* (3), 1758.

(53) Perdew, J. P.; Burke, K.; Wang, Y. Generalized Gradient Approximation for the Exchange-Correlation Hole of a Many-Electron System. *Phys. Rev. B* **1996**, *54* (23), 16533.

(54) Datta, D.; Li, J.; Shenoy, V. B. Defective Graphene as a High-Capacity Anode Material for Na- and Ca-Ion Batteries. *ACS Appl. Mater. Interfaces* **2014**, *6* (3), 1788–1795. https://doi.org/10.1021/am404788e.

(55) Mukherjee, R.; Thomas, A. V.; Datta, D.; Singh, E.; Li, J.; Eksik, O.; Shenoy, V. B.; Koratkar, N. Defect-Induced Plating of Lithium Metal within Porous Graphene Networks. *Nat. Commun.* **2014**, *5*. https://doi.org/10.1038/ncomms4710.

(56) Bader, R. F. W. Atoms in Molecules. *Acc. Chem. Res.* **1985**, *18* (1), 9–15. https://doi.org/10.1021/ar00109a003.